\documentclass[aps,prl,showpacs,groupedaddress,twocolumn]{revtex4}

\usepackage{graphicx}
\usepackage{amssymb}
\usepackage{verbatim}

\def\Mvariable#1{#1}

\begin{document}
\newcommand{\sgn}{\,{\rm sgn}\,}
\setlength{\unitlength}{1cm}

\title{Signatures of fractional Hall quasiparticles in moments of current through an antidot}

\author{Alessandro Braggio$^{1}$, Nicodemo Magnoli$^{2}$, Matteo Merlo$^{1,2}$,
and Maura Sassetti$^{1}$}
  \affiliation{
  $^{1}$Dipartimento di Fisica, INFM-LAMIA,
  Universit\`{a} di Genova, Via
  Dodecaneso 33, I-16146 Genova, Italy \\
  $^{2}$Dipartimento di Fisica, INFN,
  Universit\`{a} di Genova, Via
  Dodecaneso 33, I-16146 Genova, Italy \\}
 \date{\today}
\begin{abstract}
The statistics of tunneling current in a fractional quantum Hall sample with an antidot is studied
in the chiral Luttinger liquid picture of edge states. A comparison between Fano factor and skewness is proposed
in order to clearly distinguish the charge of the carriers in both the thermal and the shot limit. In addition, we address effects
on current moments of non-universal exponents in single-quasiparticle propagators. Positive correlations, result of propagators behaviour,
are obtained in the shot noise limit of the Fano factor, and possible experimental consequences are outlined.

\end{abstract}


\maketitle

\emph{Introduction} - The properties of quasiparticles in the fractional quantum Hall effect (FQHE) have received great attention
since Laughlin's work for the states at filling factor $\nu=1/p, \, p$ odd integer, in which gapped bulk excitations were predicted
to exist and to possess fractional charge $e^*=\nu e$ ($e=$ electron charge)~\cite{Laughlin}.
A theory of the FQHE in terms of edge states has been proposed by Wen~\cite{Wen}. This theory recovered the
fractional numbers of quasiparticles in the framework of chiral Luttinger Liquids ($\chi$LL), and indicated tunneling as an
accessible tool to probe them~\cite{kanefisher}.
A charge $e/3$ of quasiparticles in the $\nu=1/3$ state was indeed measured in shot
noise experiments with point-contact geometries and edge-edge backscattering~\cite{depicciotto}.
In addition, $\chi$LL theory predicts a universal interaction parameter equal to $\nu$.  The
resulting edge tunneling  density of states should reflect in a power-law behaviour of  $I-V $ curves with  universal exponents, e.g.
 $\nu^{-1}$ in the case of metal-edge tunneling~\cite{kanefisher}. Experiments with edge states at filling $1/3$ indeed
 proved a power-law behaviour but with an exponent
different from $3$~\cite{chang}, and deviations were observed also in the point contact geometry~\cite{depicciotto,beltram}.
The disagreement of $\chi$LL predictions with observed exponents is still not completely understood,
although several mechanisms have been put forward,
including coupling to phonons~\cite{halpros,eggert}, interaction with reservoirs~\cite{ponomarenkonagaosa}, and
edge reconstruction with smooth confining potentials~\cite{yang}.

\begin{figure}
 \includegraphics[angle=0,width=8.6 cm]{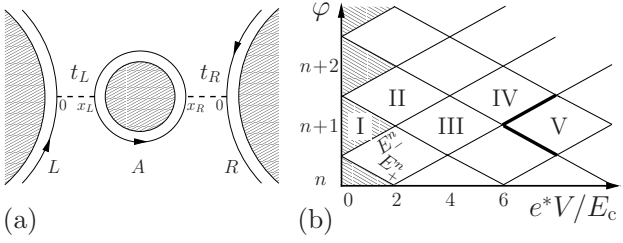}
 \caption{(a) Geometry of the system.
(b) Scheme of transport regions in the $(V,\varphi)$ plane. Roman numbers  indicate  the number of charge
states involved in the transport - hatched, blockade regions. Thin lines signal the onset
of transitions, where energies $E_\pm^n=0$ (see text). Thick lines indicate the diamond where
plasmonic excitations first enter into effect for $\nu=1/3$ (region V).  }\label{figureone}
\end{figure}

In this Letter, we aim to find signatures of fractional charge in different transport regimes, and to distinguish
 them from effects due to  quasiparticle
propagators. We consider a system consisting in a quantum
Hall sample  with an embedded antidot (Fig.~\ref{figureone}(a)) at filling factor $\nu=1/p$.
We derive unambiguous signatures of the fractional charge
in processes with different transport statistics
through a comparison of noise and skewness in the sequential regime. In addition, we find transport regimes where the Fano factor is
 sensitive to the power laws of the quasiparticle propagators and presents super-poissonian correlations.
 The peculiar behaviour driven by the quasiparticles could allow for a direct estimate of possible renormalization effects in propagators.
 The choice of the system has been motivated by recent experiments~\cite{gold} on fractional charge and statistics. These geometries
 seem indeed a promising candidate to verify experimentally our predictions.

\emph{Model} -  Edge states form at the boundaries of the sample and around the antidot (Fig.~\ref{figureone}(a));
tunneling barriers couple the antidot with both edges. The Hamiltonian reads
$H=H^0_{{L}}+H^0_{{R}}+H^0_{{A}}+H^{\rm AB}+H_R^{{T}}+H_L^{{T}}$, where the
$H^0_l$  are Wen's  Hamiltonians for the left, right and antidot edge ($l=L,R,A$), $H^{\rm AB}\propto \mathbf{j}_{A}\cdot \mathbf{A}$
describes the coupling of the antidot current
$\mathbf{j}_{A}$ with the vector potential $\mathbf{A}$, and $H^{{T}}_i$ is the tunneling between the $i=L,R$ infinite edges and the antidot.
 With $\hbar=1$, one has~\cite{Geller97}
\begin{equation} \label{hamilt}
H_l^0=\frac{v}{4\pi \nu}\int {\rm d}x \left(\partial_x\phi_l(x)\right)^2,
\end{equation}
where $v$ is the edge magnetoplasmon velocity and $\phi_l(x)$ are scalar fields satisfying
the equal-time commutation relations $[\phi_l(x),\phi_{l'}(x')]=\pm i \pi \nu \delta_{ll'}\sgn(x-x')$ whose sign depends on the
chirality.
For the antidot of length $\mathrm{L}$,  the field $\phi_A(x)$  comprises a zero-mode describing the charged excitations and a
neutral boson satisfying periodic boundary conditions, $H^0_{{A}} =E_{\rm c} n^2 +\sum_{l>0} l \epsilon a_l^\dagger a_l$.
 Here, $E_{\rm c}=\pi  \nu v/\mathrm{L}$ is the topological charge excitation energy, and
 $n$ is the excess number of elementary quasiparticles; for the neutral sector, $a,a^\dagger$ are  bosonic operators (plasmons)
 and $\epsilon= 2\pi  v/\,\mathrm{L}$ is the plasmonic excitation energy~\cite{Geller97}. The effect of $H^{\rm AB}$ is merely to
 shift the energies in $H^0_{{A}}$
  according to $E_{\rm c} n^2\to E_{\rm c} (n-\varphi)^2$, where $\varphi=\Phi/\Phi_0$
  is the Aharonov-Bohm flux $\Phi$ through the antidot measured in flux quanta $\Phi_0=hc/e$~\cite{Geller97}.

   The term $H^{{T}}=\sum_{i=L,R} t_{{i}}\psi_{{A}}^\dagger(x_{ i})\psi_{{i}}(0)  + h.c.$
represents the most relevant processes of single-quasiparticle tunneling~\cite{kanefisher,Geller97}. Here,
$\psi^\dagger_l(x)\propto e^{-i \phi_l(x)}$ are the creation operators for quasiparticles in the leads and in the antidot. Standard
commutation relations ensure a charge of quasiparticles $e^*=\nu e$~\cite{Wen}. The tunneling probability ratio between the
two barriers is tuned by an asymmetry $\eta=\vert t_R\vert^2/\vert t_L\vert^2$.
 A source-drain voltage $V$ is applied between the left and right edges, producing a backscattering tunneling
 current of quasiparticles through the antidot $I(t)=\left[I_L(t)-I_R(t)\right]/2$, with ($j=L,R$)
  \begin{equation}\label{current}
 I_{j}(t)=ie^*  \bigg[   t_{j} \psi_{{A}}^\dagger(x_{j},t)\psi_{{j}}(0,t) - h.c. \bigg].
\end{equation}

\emph{Sequential tunneling rates} - For sufficiently small tunneling as compared with temperature, transport can
be safely described within the sequential  tunneling regime~\cite{furusaki}.
 Here, the main ingredients are the incoherent tunneling rates $\Gamma_{L,R}(E)$. Their expression is well known within
 the Luttinger description of edge states with fully relaxed plasmonic excitations of the antidot~\cite{GellerLoss2,braggio}. One has
 $\Gamma_i(E)=\vert t_i\vert ^2\Gamma(E)= \vert t_i\vert^2 \sum_l w_l \gamma(E-l\epsilon)$
 where $E$ is the energy associated to the quasiparticle tunneling event and
$\gamma (x)=(\beta\omega_{\rm c}/2 \pi)^{1-g} \vert \mathbf{\Gamma}(g/2+i\beta x/2\pi)\vert^2 e^{\beta x/2}$ with $\mathbf{\Gamma}(x)$
the Euler Gamma function  and $\beta=1/k_{\rm B}T$. The factors $w_l$ are function of the plasmonic energy $\epsilon$, the interaction parameter
 $g$ and the cut-off energy $\omega_{\rm c}$~\cite{braggio}. Note that in the standard $\chi$LL theory  $g=\nu$.
  Here, we will assume $g=\nu F$ in order to describe possible renormalization effects due to coupling of the infinite
 edges with additional modes, e.g. phonons, or to edge reconstruction. The explicit value of $F$ will depend on the details of
 interaction~\cite{halpros,eggert,ponomarenkonagaosa,yang} and here we will consider it as a free parameter.
 Note also that the fractional charge $e^*$ is solely determined by $\nu$ and is thus separated from the dynamical behaviour
 governed by $g$.\\
 For $g<1$  the rates scale at low temperatures as $T^{g-1}$  at energy $E=l\epsilon$. This behaviour is reflected in the
  increase of the linear conductance maximum  $G_{\rm max} \propto T^{g-2}$ with decreasing temperature. In order to be consistent
 with the tunneling approximation we then require $G_{\rm max}\ll e^2/h$, setting a limit to the low
 temperature regime~\cite{furusaki}.

\emph{Moments} - Hereafter, we will study higher current moments
 as a tool to determine the $\chi$LL exponent~\cite{ponomarenkonagaosa} and the carrier charge, decoupling the latter from the information
 on the statistics of the transport process.
We will consider the $n$-th order normalized current cumulant~\cite{LR01},
 \begin{eqnarray}
\label{norm-moments}
k_n&=&\frac{\langle \langle I   \rangle \rangle_n }{|e^{n-1} \langle I  \rangle |}.
\end{eqnarray}
Here, $\langle I  \rangle$ is the stationary current
and $\langle\langle I \rangle\rangle_n=\lim_{\tau\to\infty}(e^*)^n
\langle\langle N_{\tau} \rangle\rangle_n /\tau$
is the $n$-th irreducible current moment given in terms  of the irreducible moments of
the  number $N_\tau$ of  charge $e^*$ particles  transmitted in the  time $\tau$.
Fano factor and normalized skewness correspond to $k_{2,3}$.\\
The statistics of a transport process is completely identified  by its cumulants $\langle\langle N_\tau \rangle\rangle_n$.
Indeed, if a process with a given statistics takes place at different filling factors with   $e^*_1=\nu_1 e$ and $e^*_2=\nu_2 e$,
then the comparison of the $n$-th order current cumulants gives direct information on the charge ratio according to
$k_n(\nu_1)/k_n(\nu_2)=(e^*_1 /e^*_2)^{n-1}$~\cite{saleurweiss}.
  We suggest to revert this approach to detect the charge fractionalization in our antidot geometry.
  To do so, we define \emph{special} the conditions in the parameter space where
 our system has the same transport statistics for different filling factors and independently from the value of
  $g$~\cite{nota}.
  Note that  a comparison of all moments would be required to identify special regimes. Here, we will adopt
  only the minimal comparison of the second and third moment that are more accessible in experiments.
  Furthermore, unlike simpler geometries  our system offers  the possibility to identify several  special  points with different
  statistics by changing external parameters.\\
The detailed analysis of $k_{2,3}$ is obtained directly from the cumulant
generating function calculated in the markovian master equation framework~\cite{nazbag} in the sequential regime.
The stationary occupation probability of a fixed number of antidot quasiparticles is
 obtained  in analogy to the electron number occupation in quantum dots~\cite{GellerLoss2}.
 Assuming a symmetric voltage drop at the barriers, the change in energy for the forward transitions $n\to n+1$, $n+1\to n$ is
 $E_\pm^n=e^*V/2\pm2 E_{\rm c}(\varphi-n-1/2)$
 respectively. The conditions $E_\pm^n=0$
grid the $(V,\varphi)$ plane into diamonds according to the scheme in Fig.~\ref{figureone}(b).

\emph{Results} - We focus at first on the few-state regime $e^* V\lesssim 2E_{\rm c}$. In regions I transport
 is suppressed; linear conductance oscillations exist
 in regions I,II with a periodicity of a flux quantum $\Phi_0$
 for any $\nu$, in accordance with gauge invariance~\cite{byeryang}.
In the same regime, an analytical treatment of $k_{2,3}$ is possible. Since
the energy spectrum is periodic in $\varphi$, we start at $n=0$ . Here
the forward tunneling rates $\Gamma_\pm^0 = \Gamma\left(E^0_\pm\right)$ dominate the dynamics and
we recover a known formula~\cite{BlanterReview,braggio} for the Fano factor
\begin{equation}\label{fano}
\frac{ k_2}{\nu}=\coth(\frac{\beta e^*V}{2})-2 \eta \frac{ \Gamma_+^0 \Gamma_-^0 f_-(e^* V)}{\Gamma_{\rm tot}^2},
\end{equation}
where $\Gamma_{\rm tot}= \Gamma_+^0 f_+(E_+^0) + \eta \Gamma_-^0 f_+(E_-^0)$ with $f_\pm(x)=1\pm e^{-\beta x}$.\\
For the skewness we find
\begin{equation}\label{skewness}
\frac{ k_3}{\nu^2}=1-6 \eta \frac{ \Gamma_+^0 \Gamma_-^0 f_+(e^* V)}{\Gamma_{\rm tot}^2}+
12 \eta^2 \frac{  {\Gamma_+^0}^2 {\Gamma_-^0}^2 f_-^2(e^* V)}{ \Gamma_{\rm tot}^4}.
\end{equation}

\begin{figure}
\begin{center}
\includegraphics[width=8.6 cm,angle=0]{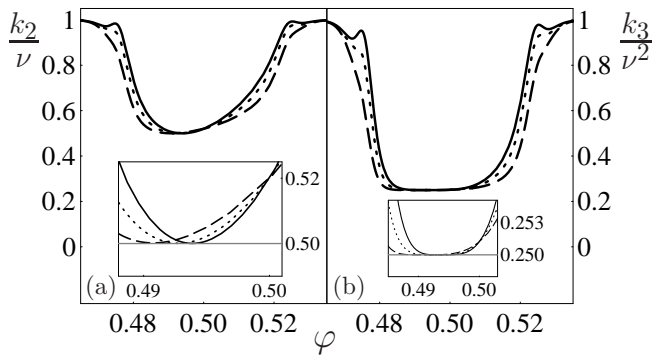}
\caption{Shot noise limit with $\eta=1.5$, $e^*V=0.1 E_{\rm c}$, $k_{\rm B}T=0.004 E_{\rm c}$ and different
$g=1/5$ (solid), $g=1/3$ (dotted), $g=1/2$ (dashed). (a) Fano factor $k_2/\nu$ and
(b) skewness $k_3/\nu^2$ vs. magnetic  flux $\varphi$. Insets: zoomed regions of
the minima, with grey lines at (a) 1/2 and (b) 1/4.}\label{figuretwo}
\end{center}
\end{figure}
We analyze now the behaviours (\ref{fano}) and (\ref{skewness}) varying the ratio $e^*V/k_{\rm B} T$.

Shot noise limit $k_{\rm B}T \ll e^* V$. In the blockade regions I  with $|\beta E_\pm^0|\gg 1$,
 one has $k_{2}=\nu$ and $k_3=\nu^2$. In this
 case the statistics of the transport process is poissonian: the transport
 through the antidot is almost completely suppressed, $I\approx 0$, and the residual
 current is generated only by a thermally activated tunneling that is completely uncorrelated. So
  in region I for a fixed value of filling factor, $k_{2,3}$
 take maximal values corresponding to a poissonian transport process, thus constituting an example of special regime.\\
We consider the two-state regime (II)  for $\beta E_\pm^0 \gg 1$.
For fractional edges $g<1$, $k_{2,3}$ have a particular functional dependence on $\varphi$.
We find that they both develop a minimum~\cite{nota2} and that
the absolute values of the minima are, respectively, $k_2^{\rm min}=\nu/2$ and $k_3^{\rm min}=\nu^2/4$. These minimal values do  not depend on $g$,
as the comparison of solid ($g=1/5$),  dotted ($g=1/3$) and dashed ($g=1/2$) curves in Fig.~\ref{figuretwo}  confirms.
For Fermi liquid edges $g=1$,
we have  $k_2=\nu(1+\eta^2)/(1+\eta)^2$ and \mbox{$k_3=\nu^2\left[1-6\eta(1+\eta^2)/(1+\eta)^4\right]$} independently from $\varphi$.
Here, $k_2$ and $k_3$  assume their minimal values $\nu/2$ and $\nu^2/4$  in the symmetric case $\eta=1$.
In this conditions we have the strongest anticorrelation that is signalled by a marked sub-poissonian statistics.\\
We can conclude that in the two-state regime, in the shot noise limit, the values of the minima for $k_{2,3}$
obtained varying $\eta,\varphi$ correspond to a special condition where the system shows
the same sub-poissonian statistics (strongest anticorrelation) for any $g\leq1$.\\
In the intermediate regime $e^* V \approx k_{\rm B}T $, $k_{2,3}$ depend more strongly on the parameter $g$, and the interplay
of two energy scales prevents the onset of special regimes.

Thermal regime $e^* V \ll k_{\rm B}T $. In this limit the Fano factor is independent from the charge
fractionalization, $k_2=2 k_{\rm B} T/e V$, reflecting the fluctuation-dissipation theorem. This is not
 true for the normalized skewness, that measures the fluctuation asymmetry induced by the current.
In this regime, the skewness opens the possibility to measure the carrier charge $e^*=\nu e$ that
 is no more addressable via the Fano factor.
Indeed, for  low voltages $V\to 0^+$  one has
\begin{equation}
\label{SkTerm}
k_3=\nu^2\left[1-3\frac{\eta}{(1+\eta)^2}\frac{1}{{\rm Cosh}^2( \beta E_{\rm c}(\varphi-1/2))}\right],
\end{equation}
that does depend on $\nu$ but not on the exponent $g$.

 \begin{figure}
\begin{center}
\includegraphics[width=8.6 cm]{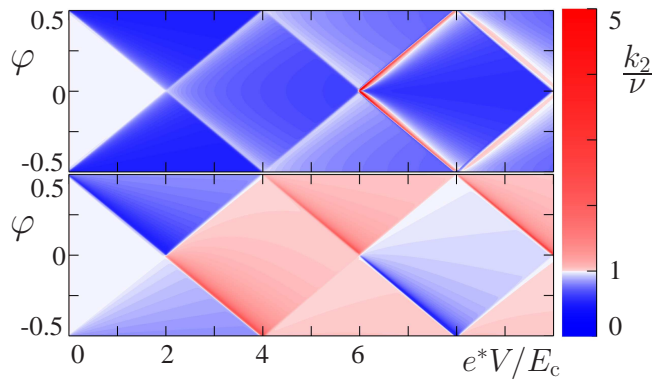}
\caption{Fano factor $k_2/\nu$ at $\nu=g=1/3$, $k_{\rm B}T=0.01 E_{\rm c}$, vs. source-drain voltage and magnetic
flux. Top panel: symmetric barriers $\eta=1$. Bottom
panel: strong asymmetry $\eta=10$. Right panel: color scale.}\label{figurethree}
\end{center}
\end{figure}

\begin{figure}
\begin{center}
\includegraphics[scale=.94]{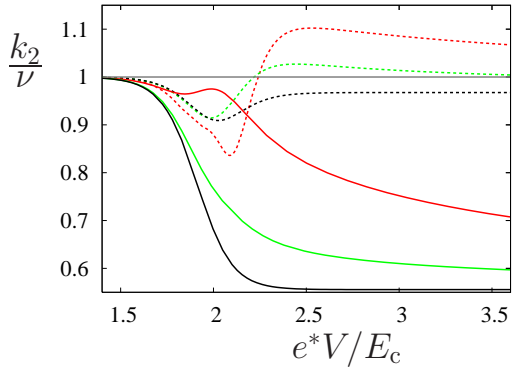}
\caption{Fano factor $k_2/\nu$ vs. $e^*V/E_{\rm c}$ at $\varphi=0, k_{\rm B}T=0.05 E_{\rm c}$. Color code:
red $g=1/3$, green $g=0.8$,  black $g=1$. Solid lines $\eta=1$, dashed lines $\eta=10$.  Grey line, poissonian limit. }\label{figurefour}
\end{center}
\end{figure}

We study now  higher voltages $e^*V > 2E_{\rm c}$ where the renormalized interaction parameter $g$ has a prominent role.
 For this purpose we consider the behaviour of the Fano factor.
Here in general a numerical approach is necessary. In Fig.~\ref{figurethree} a density plot of  $k_2$ for $\nu=g=1/3$
 as a function of magnetic field and source-drain voltage
 is shown  for different asymmetries.
 We recover that, independently from $\eta$, in region I one has $k_2=\nu$ that
 corresponds to a poissonian statistics. We will thus
 refer to the red (blue) regions, where $k_2 >\nu$ ($k_2 <\nu$), as super(sub)-poissonian noise regimes.\\
 In the three-state regime III, a comparison of the top and
 bottom panels shows that super-poissonian values are induced by high barrier asymmetry.
 The Fano factor in this regime depends on a larger set of rates $\Gamma^n_\pm=\Gamma(E_\pm^n)$, $n =0,1$, and the corresponding backward rates
 $\overline{\Gamma}_\pm^n=e^{-\beta E^n_{\pm}}\Gamma_\pm^n$.
 A tractable analytical formula can be derived under the reasonable assumption that only two  backward rates, $\overline{\Gamma}_-^0$ and
 $\overline{\Gamma}_+^1$,  survive: one has $k_2/\nu=1-2\eta\,\,\delta k_2 $, with
\begin{equation}\label{deltak}
 \delta k_2  \!=\! \frac{  {\Gamma^0_t}^2\Gamma_+^1 \Gamma_-^1 \!+\!{\Gamma^1_t}^2\Gamma_+^0 \Gamma_-^0 \!+\!
 \Gamma_-^0 \Gamma_+^1 \left(\Gamma^0_t\!-\!\Gamma^1_t\right) \left(\eta\Gamma_-^1\!-\!\Gamma_+^0\right) }
  {\left[ \eta \Mvariable{ \Gamma_-^0}
         \Mvariable{ \Gamma_t^1} + \Mvariable{ \Gamma_t^0} \left( \Mvariable{ \Gamma_+^1} + \Mvariable{ \Gamma_t^1} \right)  \right]^2}.
\end{equation}
 Here, $\Gamma^0_t=\Gamma_+^0 + \eta \overline{\Gamma}_-^0$ and $\Gamma^1_t=\eta\Gamma_-^1 +   \overline{\Gamma}_+^1  $.
 We note that in order to have super-poissonian noise a fractional  $g<1$ is necessary,
 with additional conditions on the  asymmetry. Indeed, setting
 $\eta=1$ in Eq.~(\ref{deltak}) in the  limit $\beta E_+^1,   \beta E_-^0 \gg 1$
 yields $\delta k_2>0$ for any $g$.
 On the other side, setting $g=1$ gives $\delta k_2=2\eta/(\eta^2+\eta+1)^2>0$.\\
 So it appears that positive correlations are induced by an interplay of $\eta$ and $g$.
 Figure~\ref{figurefour} shows the Fano factor as a function of $e^* V/E_{\rm c}$
 for  different $\eta$ and $g$. Here $\varphi=0$, although similar considerations apply in general.
 For $g=1$, $k_2$ remains sub-poissonian (black lines), while positive correlations appear for $g<1$ and sufficient asymmetry (color lines).

 Finally, interesting effects take place in the five-state regime (V) for $\nu=1/3$.
 Here, a superpoissonian Fano factor appears along the diamond lines for $\eta=1$ (see Fig.~\ref{figurethree}  top)
and disappears for large asymmetries  (Fig.~\ref{figurethree}  bottom).
 Detailed investigations~\cite{unpub} show that
 the collective excitations of the antidot edge  are responsible of the super-poissonian behaviour at small asymmetries.
  In this region, in fact, the tunneling process can excite the plasmonic modes of energy $\epsilon=2E_{\rm c}/\nu$ that exactly correspond to the diamond lines
  (Fig.~\ref{figureone}(b), thick curves).
  In particular, one can show~\cite{unpub} that $k_2$  shows a superpoissonian maximum as a function of $e^*V$ with a peculiar scaling law
  $k_2^{\rm max} \propto  T^{{g}-1} $ directly connected to the renormalized lead exponent.

In conclusion, we have found distinct, unambiguous signatures of fractional charge and interaction renormalization in
 high moments of tunneling current in a promising Hall-antidot geometry~\cite{gold}.
 Confirmation of such novel results appears to be within experimental reach,
 especially on account of recent accomplishments in measurement techniques applied to electron counting~\cite{skewness}.

 Financial support by the EU via Contract No. MCRTN-CT2003-504574 is  gratefully acknowledged.


\begin{thebibliography}{99}
\bibitem{nota} In the point contact geometry the tunneling limit is a
  special regime, with Poissonian statistics e.g. for both $\nu=1$ and $\nu=1/3$.
\bibitem{nota2} An analytical approximation for the minimum is given by
 $\varphi^{\rm min}=1/2-(e^* V/4E_{\rm c})(1-\eta_{\rm eff}^{\alpha(g)})/(1+\eta_{\rm eff}^{\alpha(g)})$,
with $\eta_{\rm eff}=\eta[1+O(\beta E_{\rm c})]$.
\end{thebibliography}
\end{document}